\documentclass[dvips]{article}

\def \gcm {$\mathrm{\ g\ cm^{-2}}$}
\def \ev { $\mathrm{\ eV} $}

\def \eev { $\mathrm{\ EeV}$}

\def \flxn {$\mathrm{\ GeV\ cm^{-2}\ sr^{-1}\ s^{-1}}$}
\def \deg {$^\circ$}

\def \mal {Malarg\"{u}e}

\usepackage{icrctc07}

\title{Limits to the diffuse flux of UHE tau
  neutrinos at EeV energies from the Pierre Auger Observatory}
\shorttitle{Neutrino limit from Pierre Auger Observatory}
\authors{O. Blanch Bigas$^{1,2}$, for the Pierre Auger Collaboration$^{3}$}
\shortauthors{The Pierre Auger Collaboration}
\afiliations{$^1$ LPNHE, IN2P3 - CNRS - Universit\'{e}s Paris VI et
  Paris VII, 4 place Jussieu, Paris, France\\$^2$Ministerio de Educaci\'{o}n
  y Ciencia, Spain\\$^3$ Observatorio Pierre Auger, Av. San Martin Norte 304, (5613) \mal, Mendoza, Argentina}
\email{blanch@lpnhe.in2p3.fr}

\abstract{With the Pierre Auger Observatory we have the capability of detecting
ultra-high energy neutrinos by searching for very inclined showers
with a significant electromagnetic component. In this work we discuss
the discrimination power of the instrument for earth skimming tau
neutrinos with ultra-high energies. Based on the data collected since
January 2004 an upper limit to the diffuse flux of neutrinos at\eev
energies is presented and systematic uncertainties are discussed.}

\begin{document}
\maketitle

\section{Introduction}
The nature and the production mechanism of the cosmic rays of
ultra-high energy (UHE), above $10^{19}$\ev, is still unknown. All proposed
mechanisms are expected to produce neutrinos. Classical acceleration
process of charged particles in astrophysical objects create
neutrinos through interactions with the radiation within the source
region or with the Cosmic Microwave Background (GZK neutrinos) \cite{ref1}. In
other type scenarios they arise as direct or indirect products of
supermassive particles. The $\tau$ neutrinos are heavily suppressed at
production. In the scenario of neutrino flavor oscillation and a
maximal $\Theta_{23}$ mixing, the flavor balance changes when
neutrinos reach the earth. After travelling cosmological distances,
approximately equal fluxes for each flavor are obtained
\cite{ref2}. Tau neutrinos that enter the earth just below the
horizon, the so-called skimming neutrinos, may undergo a
charged-current interaction to produce a $\tau$. When the interaction
happens close to the surface a $\tau$ can exit the earth and its decay
in the atmosphere can produce an Extended Air Shower (EAS) detectable
with the Pierre Auger Observatory \cite{ref3}. In the\eev range, this
channel has been shown to increase the prospect of detecting UHE
neutrinos \cite{ref4}.

\section{Search for neutrinos}
UHE particles interacting in the atmosphere give rise to EAS with the
electromagnetic component reaching its maximal development after a
depth of the order of 1000\gcm$\mathrm{\ }$and extinguishing gradually
within the next 1000\gcm. After a couple of vertical
atmospheric depths only the muons survive. As a consequence very
inclined showers induced by nuclei (or possibly photons) in the upper
atmosphere reach the ground as a thin and flat front of hard muons. On the
contrary, if a shower begins development deep in the atmosphere (a tau
decay) its electromagnetic component can reach the ground and give a
distinct broad signal. Therefore, the detection of very inclined showers with a
significant electromagnetic component are a clear indication for UHE
neutrinos.

\begin{figure}
\begin{center}
\includegraphics [width=0.48\textwidth]{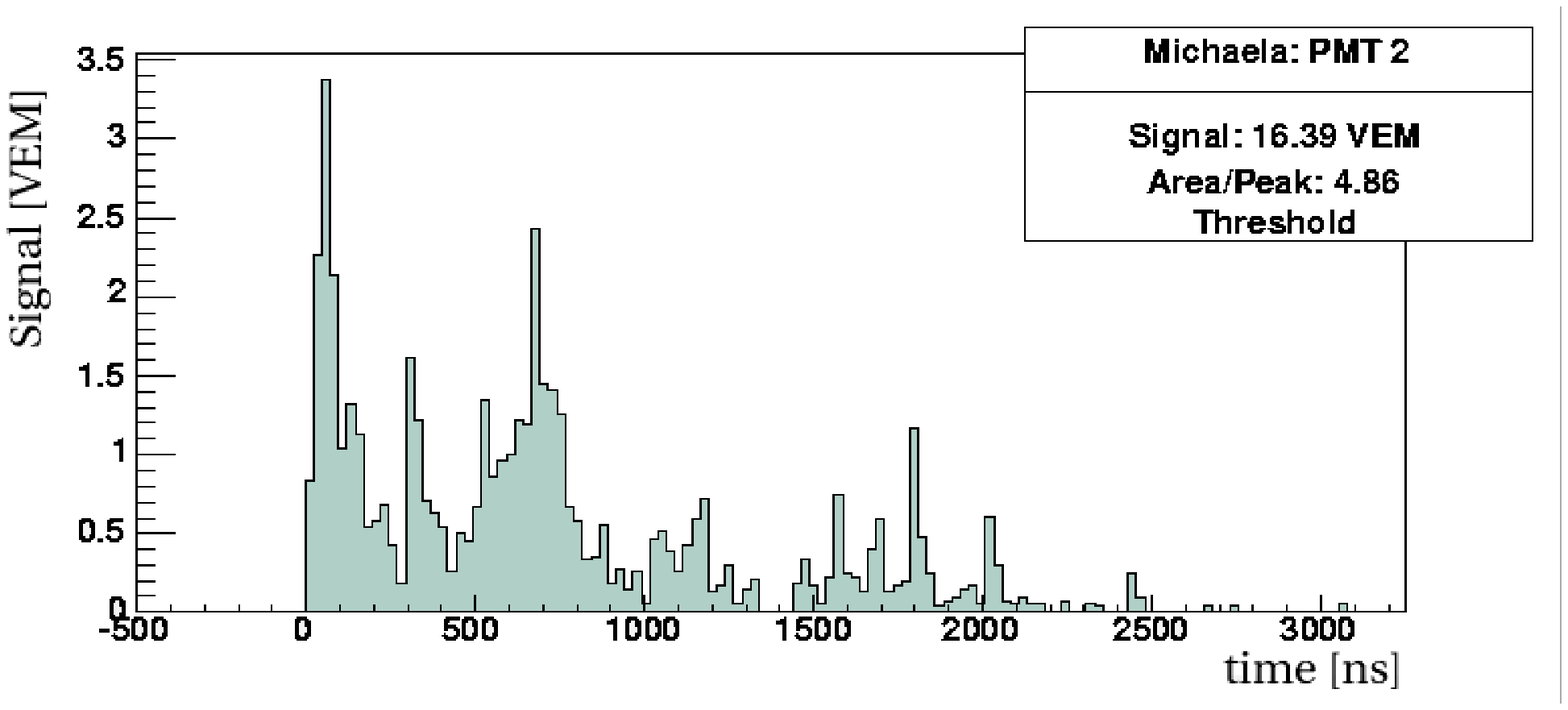}
\includegraphics [width=0.48\textwidth]{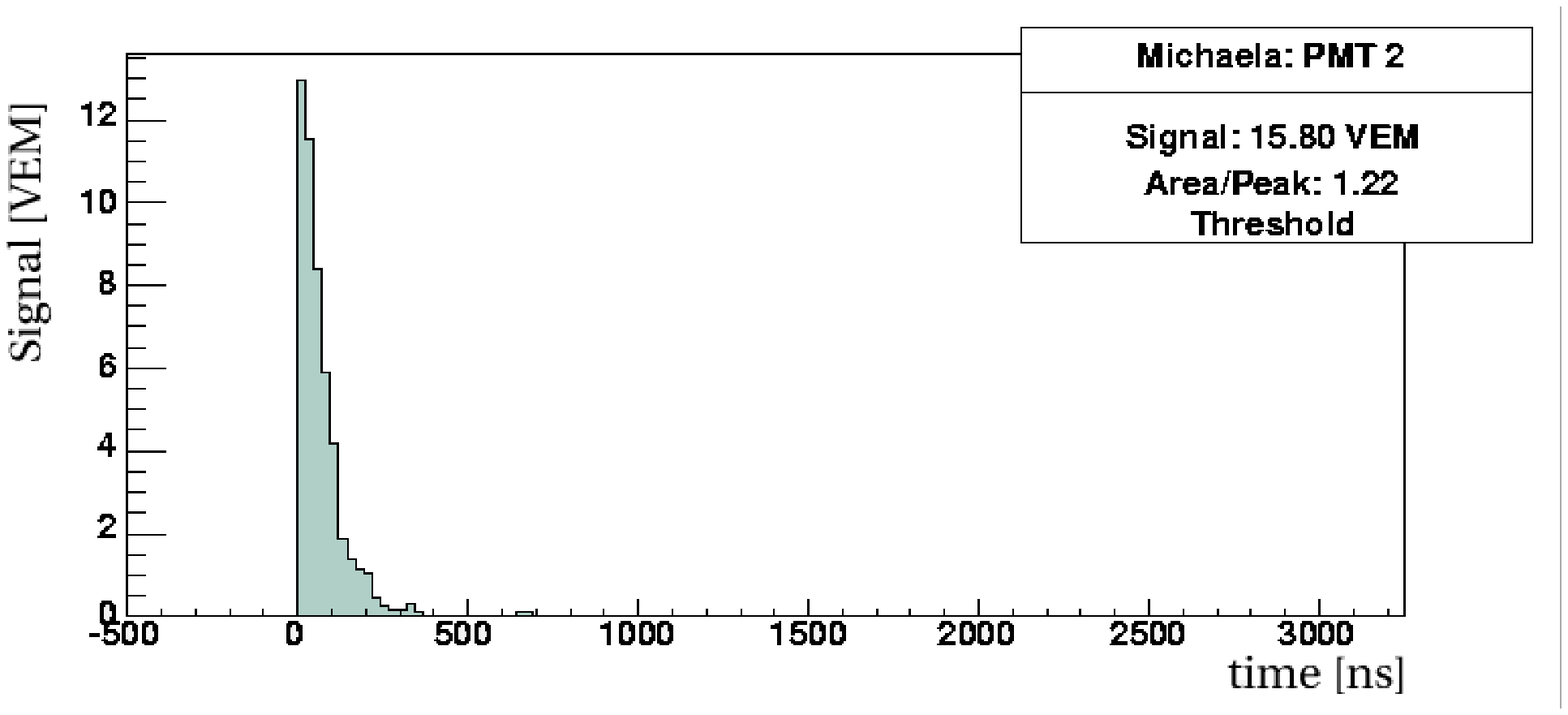}
\end{center}
\caption{FADC traces from a station of two different
   real showers after subtraction of baseline and calibration. Top:
   moderately inclined (40\deg); bottom: quasi-horizontal
   (80\deg).}\label{fig1}
\end{figure}

The signal in each station of the surface detector is digitised
using FADCs, allowing us to unambiguously distinguish the narrow
signals from the broad ones and thus to discriminate stations
with and without electromagnetic component (figure \ref{fig1}). We tag
the stations for which the main segment of its FADC trace has 13 or
more neighbour bins over the threshold of 0.2 VEM \cite{ref5} and
the area over peak ratio \cite{ref6} is larger than 1.4. The
event is selected if the tagged stations fulfil the trigger condition
and they contain most of the signal. After this selection, an almost pure
sample of young showers is isolated.

\begin{figure*}
\begin{center}
\includegraphics [width=0.96\textwidth]{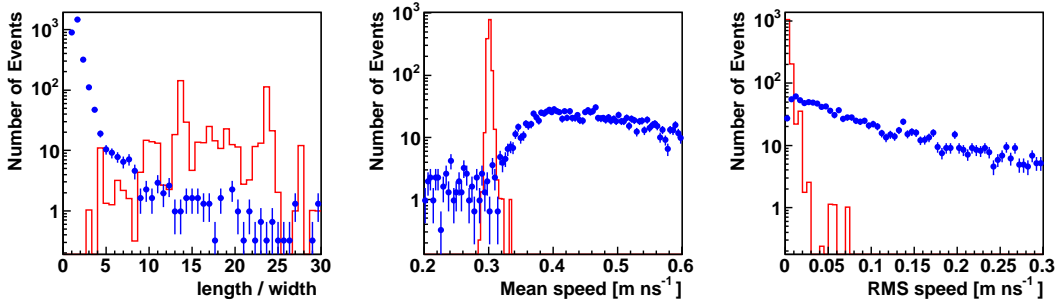}
\end{center}
\caption{Distribution of discriminating variables
  for neutrinos with an E$^{-2}$ flux (histogram) and real events passing the
  ``young shower'' selection (points). Left: length/width ratio;
  middle: average of the speed between pairs of stations; right:
  r.m.s. of the speeds.}\label{fig2}
\end{figure*}

The next step uses the footprint of local stations included in
the global trigger to select very inclined showers. First a tensor is
built using the station signals and the ground positions (in analogy
to the inertia tensor) and the corresponding major and minor axes are
used to define a ``length'' and a ``width''. Then, for each pair (i,j)
of tanks, a ``ground speed'' is defined as $d_{i,j}/|\Delta t_{i,j}|$,
where $d_{i,j}$ is the distance between them (projected onto the major
axis) and $|\Delta t_{i,j}|$ is the difference between the start times
of their signals. Horizontal showers have an elongated shaped (large
value of length/width) and they have ground speeds tightly
concentrated around the speed of light. In figure \ref{fig2}, we
show the distributions of these discriminating variables for real
events and simulated tau showers. The following cuts are applied:
length/width $>$ 5, average speed $\in$ (0.29,0.31) m ns$^{-1}$ and
r.m.s.(speed) $<$ 0.08 m ns$^{-1}$. We keep about 80 $\%$ of the
$\tau$ showers that trigger the surface detector. The final sample is
expected to be free of background.

\section{Acceptance and neutrino limit}
Both the criteria to identify neutrino induced showers and the calculation
of the $\nu_{\tau}$ acceptance are based on Monte Carlo
techniques. The former uses the simulation of the shower development
in the atmosphere as well as the detector response. The latter needs
the simulation of the interactions that happen while the neutrino
crosses the earth \cite{ref7}.

The total acceptance collected from January 2004 until December 2006 with the
Pierre Auger Observatory is the time integration of the
instantaneous aperture. 

$$Acc(E_{\nu}) = 
 \int_{0}^{E_{\nu}}dE_{\tau}\int_{0}^{\infty}dh_{c}
 \left(\frac{d^{2}N_{\tau}}{dE_{\tau}dh_{c}} Acc_{\tau}\right)$$ 
$$Acc_{\tau}(E_{\tau},h_{c}) = \hspace{50mm}$$
$$ = \int_{T} dt \int_{A} dxdy \, I_{eff}(E_{\tau},h_{c},x,y,A_{Conf}(t))$$

where $dN_{\tau}/dE_{\tau}dh_{c}$ is the flux of emerging
$\tau$s and $I_{eff}$ the probability to identify a $\tau$. It depends
on the energy of the $\tau$ ($E_{\tau}$), the altitude of the shower
center defined 10 km after the decay point ($h_{c}$)
\cite{ref4}, the instantaneous configuration of the detector
($A_{Conf}(t)$), and the relative position of the shower footprint in
the array ($x,y$).

The $Acc(E_{\nu})$ is computed by Monte Carlo in two independent
steps. First, the integral on time and area are performed using the
simulations of the EAS and the detector, allowing us to account for
the time evolution of the detector. The second step computes the
integral on $h_{c}$ and $E_{\tau}$ by adding
$Acc_{\tau}(E_{\tau},h_{c})$ for all emerging $\tau$, given by the
simulation of the earth interactions. The statistical precision due to
the statistic of the Monte Carlo simulation is at a few percent level.

The Monte Carlo simulations use several physical magnitudes that have
not been experimentally measured at the relevant energy range,
namely the $\nu$ cross-section, the $\tau$ energy losses and the $\tau$
polarisation. We estimate the uncertainty in the acceptance due to the
first two to be 15$\%$ and 40$\%$ respectively, based on Particle
Distribution Function (PDF) uncertianties. The two polarizations give
30$\%$ difference in acceptance. We take it as the corresponding
uncertainty. The relevant range for PDFs includes combinations of x
and Q$^{2}$ where no experimental data exist. Different extrapolations
to low x and high Q$^{2}$ would lead to a wide range of values for the
$\nu$ cross-section as well as the $\tau$ energy losses. The
uncertainties on the low x regime as well as possible large $\nu$
cross-sections have not been added on the quoted systematics.

We also took into account uncertainties coming from neglecting the
actual topography around the site of the Pierre Auger Observatory
(18$\%$). We are confident on the simulations of the
interactions undergoing in the earth at 5 $\%$ level. And we quote a 25
$\%$ systematic uncertainty due to Monte Carlo simulations of the
EAS and the detector.

\begin{figure}
\begin{center}
\includegraphics [width=0.48\textwidth]{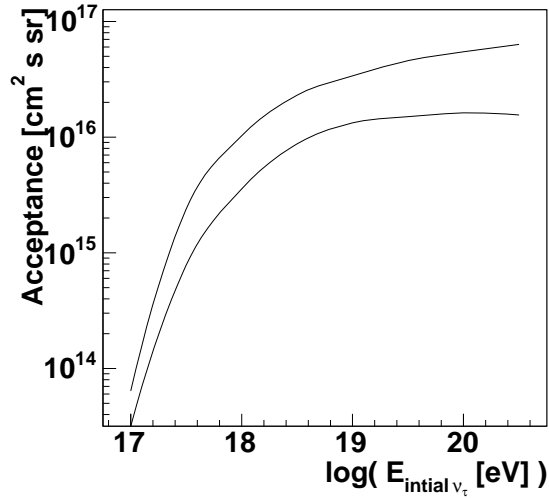}
\end{center}
\caption{The Pierre Auger Observatory acceptance from January 2004
  until December 2006 for the most and least favourable scenarios of
  the systematics.}\label{fig3}
\end{figure}

\begin{figure}
\begin{center}
\includegraphics [width=0.48\textwidth]{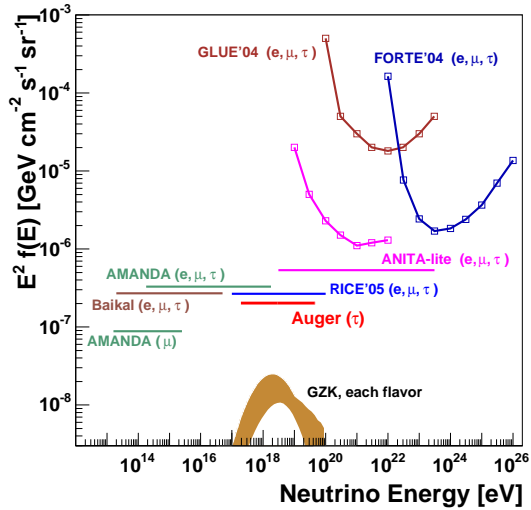}
\end{center}
\caption{Limit at 90$\%$ C.L. to an E$^{-2}$ diffuse flux of
      $\nu_{\tau}$ at\eev$\mathrm{\ }$energies from the Pierre Auger
      Observatory. Limits from other experiments
      \cite{ref10,ref11,ref12,ref13,ref14,ref15,ref16} as well as
      fluxes for GZK $\nu$ \cite{ref17,ref18} are also shown. For each
      experiment, the flavors to which is sensitive are stated.}\label{fig4}
\end{figure}

Data from January 2004 until December 2006, which equate to about 1
year from the completed surface detector, have been analysed. In
figure \ref{fig3}, we show the collected acceptance on the
analysed period, for the most and least favourable scenarios of the
systematics. Over that period, there is not a single event that fulfils
the selection criteria. Based on that, the Pierre Auger Observatory data
can be used to put a limit for an injected spectrum $K \cdot \Phi(E)$ with
a known shape. For an E$^{-2}$ incident spectrum of diffuse
$\nu_{\tau}$, the 90$\%$ CL limit is
$E_{\nu}^{2} \cdot dN_{\nu{\tau}}/dE_{\nu} < 1.5_{-0.8}^{+0.5}$
$10^{-7}$ \flxn, where the uncertainties
come from the systematics. The central value is
computed using the $\nu$ cross-section from \cite{ref8},
the energy losses from \cite{ref9} and an uniform random
distribution for the tau polarisation. The bound is drawn for the
energy range 2 10$^{17}$-5 10$^{19}$\ev$\mathrm{\ }$over which 90$\%$ of the
events are expected. In figure \ref{fig4}, we show the limit from the
Pierre Auger Observatory in the most pessimistic scenario of
systematic uncertainties. It improves by a factor $\sim$3 in the most
optimistic one. Limits from other experiments are also shown assuming a
$1:1:1$ balance among flavors due to the oscillations.

\section{Summary and Prospects}
The dataset from January 2004 until December 2006, collected by the
Pierre Auger Observatory, is used to present upper limits on the
diffuse incident $\nu_{\tau}$ flux. The skimming technique is flavor
sensitive and together with the configuration of the surface detector
gives the best sensitivity around few\eev, which is the most relevant
energy to explore GZK neutrinos. The limit is still considerably higher than
GZK neutrino predictions. Neutrinos that interact in the atmosphere
can also be distinguished from nucleon showers \cite{ref7}. Hence, the
Pierre Auger Observatory can explore UHE $\nu$s with two techniques
that depend differently on $\nu$ properties like flavour or
cross-section. The Pierre Auger Observatory will keep taking data for
about 20 years over which the bound will improve by over an order of
magnitude if no neutrino candidate is found. 



\end{document}